\documentclass[10pt,prl,twocolumn,showpacs,aps,floats,floatfix,superscriptaddress]{revtex4}

\usepackage{epsfig}
\begin{document}

\newcommand{\avk}{\langle k \rangle}
\newcommand{\fluck}{\langle k^2 \rangle}

\title{Invasion threshold in heterogeneous metapopulation networks}

\author{Vittoria Colizza}\affiliation{Complex Networks Lagrange Laboratory, Institute for
  Scientific Interchange, Torino 10133, Italy}
\author{Alessandro Vespignani}\affiliation{Complex Systems Group,
School of Informatics, Indiana University, Bloomington IN 47406}

\date{\today}

\widetext

\begin{abstract}
We study the dynamics of epidemic and reaction-diffusion
processes in metapopulation models with heterogeneous connectivity
pattern. In SIR-like processes, along with the standard local
epidemic threshold, the system exhibits a global invasion threshold. We
provide an explicit expression of the threshold that sets a critical 
value of the diffusion/mobility rate below which the epidemic is 
not able to spread to a macroscopic fraction of subpopulations. 
The invasion threshold is found to be affected by the topological 
fluctuations of the metapopulation network. The presented results 
provide a general framework for the understanding of the 
effect of travel restrictions in epidemic containment.  
\end{abstract}

\pacs{89.75.-k, -87.23.Ge, 05.70Ln}

\maketitle

The role of  heterogeneity has been acknowledged as a 
central question in the study of population biology of infectious 
diseases~\cite{May:1992,Hethcote:1978,May:1984}
and revamped recently with the evidence that a large number of real
world networks exhibit complex topological
properties~\cite{Albert:2002,mendesbook,Newman:2003}. 
These features, often mathematically encoded in a
heavy-tailed probability distribution $P(k)$ 
that any given node has degree $k$, were shown to
affect the system evolution altering 
the threshold behavior and the associated dynamical phase transition
\cite{pv01a,lloyd01,Barthelemy:2005}. 
These studies have mainly focused on networked systems where each node 
 corresponds to a single individual and only recently the study of
the impact of heterogeneous topologies on 
bosonic systems, where nodes can be occupied by any number of 
particles, has been initiated~\cite{Colizza:2007b}. Examples are provided by 
reaction-diffusion systems used to model a wide range of phenomena 
in chemistry and physics~\cite{Kampen:1981}, and metapopulation 
epidemic models~\cite{Hethcote:1978,May:1984,Keeling:2002,Lloyd:1996,Grenfell:1997,Grenfell:1998,Ferguson:2003} 
where particles
represent people moving across different subpopulations (nodes) such as city or
urban areas, and the reaction processes account for the local
infection dynamics.

Here we analyze epidemic metapopulation models characterized by 
an infection dynamics within each node (or subpopulation) 
that follows
a Susceptible-Infected-Removed (SIR) model. The mobility rate $p$ of  
individuals defines the coupling process
among the subpopulations. 
In the real world, the networks representing the mobility pattern of 
individuals among different subpopulations are 
in many cases highly
heterogeneous~\cite{Chowell:2003,transims,Barrat:2004,Guimera:2005,DeMontis:2007}. 
For this reason, the connectivity pattern of the
metapopulation network is described as a random graph with arbitrary
degree distribution $P(k)$. By using a mechanistic approach it is
possible to show that along with the usual epidemic threshold 
condition $R_0>1$ on the basic reproductive number,
the system exhibits a 
global invasion threshold setting the condition for the 
infection of a macroscopic
fraction of the metapopulation system~\cite{Ball:1997,Cross:2005}. 
The threshold condition
on  $R_0$ ensures the local outbreak at the 
subpopulation level~\cite{May:1992,Colizza:2007b}, whereas
the explicit expression obtained for global invasion threshold $R_*>1$ 
provides a critical value for
the diffusion rate $p$ below which the epidemic cannot propagate
to a relevant fraction of subpopulations.
We find that the 
global invasion threshold is affected by the topological fluctuations of the
underlying network. The larger the network
heterogeneity and the smaller is the value of the critical diffusion rate
above which the epidemic may globally invade the metapopulation
system. The present results can be generalized to more realistic
diffusion and mobility schemes and provide a framework for the
analysis of realistic metapopulation epidemic models~\cite{Grais:2003,Hufnagel:2004,Colizza:2006a,Colizza:2007a,Riley:2007}. 

A simplified mechanistic (i.e. microscopic in the epidemic
terminology) approach to the metapopulation spreading of
infectious diseases uses a
markovian assumption in which at each time step the
movement of individuals is given in terms of a matrix $d_{ij}$ that
expresses the probability that an individual in the subpopulation $i$
is traveling to the subpopulation $j$. Several modeling approaches to
the large scale spreading of infectious diseases~\cite{Rvachev:1985,Grais:2003,
Hufnagel:2004,Colizza:2006a,Colizza:2007a}
use this mobility process based on transportation
networks combined with the local evolution of the disease.  
The markovian character lies in the assumption that at each 
time step the same traveling probability
applies to all individuals in the subpopulation without having memory
of their origin. This mobility scheme coupled with an infection dynamics
at the local level can be generally viewed as
equivalent to classic reaction diffusion processes with no constraint
on the occupation numbers
$N_i$ of each subpopulation. The total
population of the metapopulation system is $N=\sum_i N_i$ and each
individual diffuses along the edges with a diffusion coefficient
$d_{ij}$ that depends on the node degree, subpopulation size and/or the
mobility matrix. The metapopulation system
is therefore composed of a network substrate connecting nodes~--~each
corresponding to a subpopulation~--~over which individuals diffuse. 
We consider that each node $i$ is connected to other $k_i$
nodes according to its degree resulting in a network with degree 
distribution $P(k)$ and
distribution moments $\langle k^\alpha\rangle=\sum_k k^\alpha P(k)$.

In the following, as a simplified diffusion process we assume 
that the mobility is equivalent to a diffusion rate along any given link of a 
node with degree $k$ simply equal to $d_{kk'}=p/k$.
This is obviously not the
case in a wide range of  real systems where 
the extreme heterogeneity of traffic is
well documented and more realistic processes will be 
considered elsewhere. This simple process however
automatically generates a stationary distribution of occupation numbers that is
better described by grouping subpopulations according to their degree $k$ 
\begin{equation}
N_k = \frac{k}{\langle k \rangle} \bar{N},
\end{equation}
where $\bar{N}$ is the average subpopulation size.

In each subpopulation $j$ the disease follows
an SIR model and
the total number of individuals 
is partitioned in the compartments 
$S_j(t)$, $I_j(t)$ and $R_j(t)$, denoting 
the number of susceptible, infected and removed individuals at
time $t$, respectively. The infection dynamics proceeds as follows. 
Each susceptible individual has a
transition rate to the infected state expressed as $\beta I_j/N_j$,
where $\beta$ is the disease transmissibility rate and $I_j/N_j$ is
the force of infection in the homogeneous mixing assumption. Analogously,
each infected individual enters the removed compartment according to
the recovery rate $\mu$. The basic SIR rules thus define a reaction
scheme of the type $S+I\to 2I$ and $I\to R$, that  conserves
the number of individuals. The SIR epidemic model is characterized by the 
reproductive number $R_0=\beta/\mu$ that defines the average number of
infectious individuals generated  by one infected individual
in a fully susceptible population. The epidemic is able to 
generate a number of infected
individuals larger than those who recover only if $R_0>1$, yielding
the classic result for the epidemic threshold~\cite{May:1992}.
If the spreading rate is not large enough  to allow a reproductive number
larger than one (i.e. $\beta>\mu$), the epidemic outbreak will quickly
die out. This result is valid at the level of each subpopulation
and holds also at the metapopulation level where
$R_0>1$ is a necessary condition to have the growth of the 
epidemic~\cite{Colizza:2007b}. 

The intuitive result on the subpopulation epidemic threshold
however does not 
take into account the effects due to the finite size of
subpopulations, the discrete nature of individuals and
the stochastic nature of the reaction and diffusion processes. These 
effects have been shown to have a crucial
role in the problem of resurgent epidemics, extinction and
eradication~\cite{Ball:1997,Cross:2005,Watts:2005,Vazquez:2007}.
Also in the present framework indeed
each subpopulation may or may not transmit the infection to 
a neighboring subpopulation  upon the condition that 
at least one infected individual is moving onto the non-infected 
subpopulations during the epidemic outbreak. 
Given an SIR model with $R_0>1$, the total
number of infected individuals generated within a subpopulation and
the mobility rate must be large enough 
to ensure the seeding of other subpopulations before the end of the 
local outbreak~\cite{Ball:1997,Cross:2005}.

As a simple example of this
effect let us consider a metapopulation system in which the initial 
condition is provided by a single infection in a subpopulation with
degree $k$ and $N_k$ individuals, given  $R_0>1$. 
In the case of a macroscopic outbreak in a closed population the total number
of infected individuals during the outbreak evolution will be
equal to $\alpha N_{k}$ where  $\alpha$ depends on the specific
disease model and parameter values used. Each infected individual
stays in the infectious state for an average time $\mu^{-1}$ equal to 
the inverse of the recovery rate, during which it can travel to 
the neighboring subpopulation of degree $k'$ with rate $d_{kk'}$.
We can therefore consider that on average the
number of new seeds that may appear into a connected
subpopulation of degree $k'$ during the
duration of the local outbreak is given by 
\begin{equation}
\lambda_{kk'}=d_{kk'}\frac{\alpha N_k}{\mu}.
\label{eq:Nk}
\end{equation}
In this perspective we can provide a characterization of 
the invasion dynamics at the level of the subpopulations, translating
epidemiological and demographic parameters into Levins-type metapopulation
parameters of extinction and invasion rate.
Let us define $D^0_k$ as the number of \emph{diseased} 
subpopulation of degree $k$ at generation $0$, i.e. those which
are experiencing an outbreak at the beginning of the process. Each
infected subpopulation  will seed~--~during the course of the outbreak~--~the 
infection in neighboring subpopulations defining the set $D^1_k$ of infected
subpopulations at  generation  1, and so on.
This corresponds to a basic branching
process~\cite{Harris:1990,Ball:1997,Vazquez:2006} where the $n-$th generation of infected subpopulations of degree $k$ 
is denoted $D^n_k$.  

In order to describe the early stage of the subpopulations invasion
dynamics we assume that the number of
subpopulations affected by an outbreak (with $R_0>1$) is small and we can 
therefore study the
evolution of the number of diseased subpopulations by using a
branching process approximation
relating $D^n_k$ with $D^{n-1}_k$. Let us consider a
metapopulation network with degree distribution $P(k)$ and $V$
subpopulations and write the
number of subpopulations of degree $k$ invaded at the generation $n$
as:
\begin{equation} 
D_k^n = \sum_{k'}D_{k'}^{n-1} (k'-1) \left[1-\left(\frac{1}{R_0}\right)^{\lambda_{k'k}}\right] P(k|k')\left(1-\frac{D_k^{n-1}}{V_k}\right).
\label{poptree-het}
\end{equation}
This equation assumes that each infected subpopulation of degree $k'$
of the $(n-1)-$th generation, 
$D_{k'}^{n-1}$, will seed the
infection in a number $(k'-1)$ of subpopulations corresponding to 
the number of neighboring subpopulations
$k'$ minus the one which originally transmitted the infection, the probability $P(k|k')$ that each of the $k'-1$ not yet infected neighboring 
subpopulations has degree $k$, and the probability to observe an
outbreak in the seeded subpopulation
i.e. $(1-R_0^{-\lambda_{kk}})$~\cite{Bailey:1975}. The last
factor stems from the probability of extinction $P_{ext}=1/R_0$ of an epidemic
seeded with a single infectious individual~\cite{Bailey:1975}.
In order to obtain an
explicit result we will consider in the following that $R_0-1\ll 1$,
thus assuming that the system is found to be
 very close to the epidemic threshold. 
In this limit we can approximate the outbreak probability as
$1-R_0^{-\lambda_{k'k}}
\simeq \lambda_{k'k}(R_0-1)$.
The case of homogeneous
diffusion of individuals $d_k=p/k$ with the stationary solution of
eq.~(\ref{eq:Nk}) for the subpopulation size yields 
$\lambda_{k'k}= p \bar{N}\alpha\mu^{-1}/\avk$. In addition, we assume that
at the early stage of the epidemic
$D_k^{n-1}/V_k \ll 1$, and
we consider the case of uncorrelated
networks in which the conditional
probability does not depend on the originating node, i.e. 
 $P(k|k')= kP(k)/\langle
k\rangle$~\cite{mendesbook}, 
obtaining 
\begin{equation} 
D_k^n =(R_0-1)\frac{kP(k)}{\langle k\rangle^2}
\frac{p\bar{N}\alpha}{\mu} \sum_{k'}D_{k'}^{n-1} (k'-1).
\end{equation}
By defining $\Theta^n= \sum_{k'}D_{k'}^{n} (k'-1)$ the 
last expression can be conveniently written in the iterative form
\begin{equation} 
\Theta^n =(R_0-1)\frac{\langle k^2 \rangle - \langle k \rangle}{\langle k \rangle^2}
\frac{p\bar{N}\alpha}{\mu} \Theta^{n-1},
\end{equation}
that allows the increasing of infected subpopulations and a global
epidemic in the metapopulation process only if 
\begin{equation} 
R_*=(R_0-1)\frac{\langle k^2\rangle - \langle k \rangle}{\langle k\rangle^2}
\frac{p\bar{N}\alpha}{\mu} >1,
\label{eq:R_*}
\end{equation}
defining the  {\em global invasion threshold} of the metapopulation system.
In other words, $R_*$ is the analogous of the basic reproductive number
at the subpopulations level and is a crucial indicator in assessing the
behavior of epidemics in metapopulation  models. 
Its expression indeed contains the probability of generating an outbreak
in a neighbor subpopulation by means of mobility processes,
$(R_0-1)p \bar{N}\alpha /(\mu \langle k \rangle)$ for $R_0-1\ll 1$,
 times the factor $\langle k^2\rangle/\langle k \rangle -1$
which also appears in the threshold conditions characterizing 
phase transitions on complex networks~\cite{pv01a,Callaway:2000,Cohen:2000,Vazquez:2006}.
The explicit form of Eq.~(\ref{eq:R_*})
can be used to find the minimum mobility rate ensuring that on
average each subpopulation can
seed more than one neighboring subpopulations. 
The constant $\alpha$ is larger than zero for any
$R_0>1$, and in the SIR case for $R_0$ close to 1 it can be approximated 
by $\alpha\simeq 2(R_0-1)/R_0^2$~\cite{Bailey:1975},
yielding for the SIR model a critical mobility
value $p_c$ below which the epidemics cannot invade the metapopulation
system given by the equation
\begin{equation} 
p_c\bar{N}= \frac{\langle k\rangle^2}{\langle k^2\rangle - \langle k \rangle}
\frac{\mu R_0^2}{2(R_0-1)^2},
\label{eq:glth1}
\end{equation}
The above condition readily tells us that the closer to the epidemic
threshold is the single subpopulation outbreak and the larger it has
to be the mobility rate in order to sustain the global spread into 
the metapopulation model. It is
important to stress that when $R_0$ increases, the small $(R_0-1)$
expansions are no longer valid and the invasion threshold is obtained
only in the form of a complicate implicit expression. 

\begin{figure}[t]
\begin{center}
\includegraphics[width=7cm]{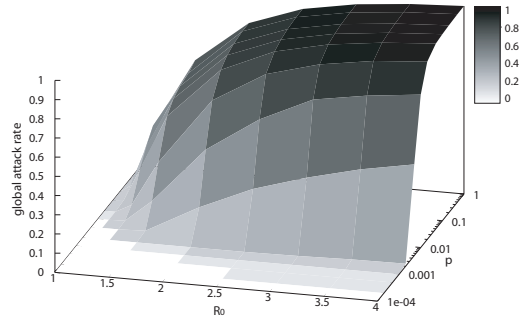}
\end{center}
\vspace{-0.5cm}
\caption{Phase diagram of the metapopulation system.
The final epidemic size is shown as a function of the local threshold $R_0$ 
and of the diffusion probability $p$.}
\label{fig:glth1}
\end{figure}
In addition, eq.~(\ref{eq:glth1}) shows the dependance of the critical
mobility on the topological fluctuations of the mobility network.
The ratio
$\langle k \rangle^2/\langle k(k-1)\rangle$ is extremely
small in heavy-tailed networks and it is vanishing in the limit of
infinite network size. This implies that the heterogeneity of the
metapopulation network is favoring the global spread of epidemics by
lowering the global spreading threshold. In other words, the
topological fluctuations favor the subpopulation invasion and
suppress the phase transition in the infinite size limit. 
This finding provides a theoretical framework and rationale for the
evidence concerning the inefficacy of travel restrictions in the
containment of global
epidemics~\cite{Hollingsworth:2006,Colizza:2007a}. 
The simple plug in of the actual numbers
for modern transportation networks, the population sizes and realistic
disease parameters in the 
expression~(\ref{eq:glth1}) indicates that a reduction of one order of
magnitude of the mobility is not enough to bring the system below the
invasion threshold. While more complicate mobility schemes should be
considered for a precise calculation, this result is setting the
framework for the understanding of mobility effects in the spreading
and containment of infectious diseases.  

In order to support the previous analytical finding we have performed
an extensive set of Monte Carlo numerical simulations of the 
metapopulation system. 
The substrate network is given by an uncorrelated complex network with
$P(k)\sim k^{-2.1}$ generated with the uncorrelated configuration
model~\cite{Catanzaro:2005}
to avoid inherent structural correlations. Network sizes of
$V=10^4$ and $10^5$ nodes have been considered. 
The dynamics proceeds in parallel and considers
discrete time steps representing the unitary time scale $\tau$ of the
process. The reaction and diffusion rates are therefore converted into
probabilities and at each time step in each subpopulation $j$ a 
susceptible individual is turned into
an infectious with probability $1-(1-\frac{\beta}{N_j}\tau)^{I_j}$ 
and each infectious individual is subject to the
recovery process and becomes recovered with probability $\mu\tau$.
The mobility is modeled assuming a diffusion probability for each individual
along each link of the subpopulation of
the form $d_{kk'}=p/k$. 

\begin{figure}[ht]
\begin{center}
\vspace{0.6cm}
\includegraphics[width=7cm]{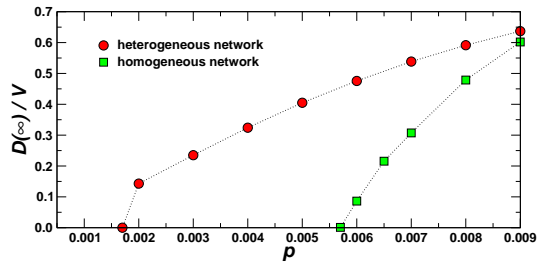}
\end{center}
\vspace{-0.5cm}
\caption{Effect of the network heterogeneity  on 
the global epidemic threshold. The final fraction of diseased 
subpopulations $D(\infty)/V$ at the end of the global epidemic 
is shown as a function of the mobility rate $p$ in a homogeneous and a
heterogeneous network.}
\label{fig:glth2}
\end{figure}
A complete analysis of the system phase
diagram is obtained by analyzing the behavior of the global
attack rate $R(\infty)/N$, defined as the total fraction of cases in the
metapopulation system
at the end of the epidemic, as a function of
both $R_0$ and $p$. Figure~\ref{fig:glth1} reports the 
global attack rate surface in the $p$-$R_0$ space and
 clearly shows the effect of different couplings
as expressed by the value of $p$ in reducing the final size of the epidemic
at a given fixed value of $R_0$. The smaller the value of $R_0$, the higher 
the coupling needs to be in order for the virus to successfully invade
a finite fraction of the subpopulations, in agreement with the
analytic result of eq.~(\ref{eq:glth1}). This provides a clear
illustration of the varying global invasion threshold as a function of
the reproductive rate $R_0$. Furthermore, 
it is possible to study the effect of the heterogeneity
of the metapopulation structure on the global epidemic threshold.  
Figure~\ref{fig:glth2} shows the results obtained by comparing two random
metapopulations networks, one with poissonian degree distribution 
(homogeneous network) and
one with heavy-tailed ($P(k)\sim k^{-2.1}$) degree distribution (heterogeneous
neywork). 
Despite the two models have the same average degree, disease 
parameters, the fluctuations of the power-law network
increase the value of $R_*$ thus lowering the critical value of the mobility.

The present analysis provides  insights in setting a framework for the
analysis of large scale spreading of epidemics in realistic mobility
networks. Furthermore,
these results open the path to future  work aimed at 
analyzing refined metapopulation
infection models.
\begin{acknowledgments}
  A. V. is partially funded by the NSF IIS-0513650. V.C. and
 A.V. are partially funded by  the CRT foundation.
\end{acknowledgments}

\end{document}